\documentclass[aps,prb,reprint,,citeautoscript,superscriptaddress,showpacs]{revtex4-1}

\usepackage{graphicx,epsfig,float}
\usepackage{amssymb} 
\usepackage{amsmath}

\begin{document}
\title{Commensurate-incommensurate phase transition and a network of domain walls in bilayer graphene with a biaxially stretched layer}

\author{Irina V. Lebedeva}
\email{liv\_ira@hotmail.com}
\affiliation{Nano-Bio Spectroscopy Group and ETSF, Universidad del Pa\'is Vasco, CFM CSIC-UPV/EHU, 20018 San Sebasti\'an, Spain}
\altaffiliation[Present address: ]{CIC nanoGUNE,
San Sebasti\'an 20018, Spain}
\author{Andrey M. Popov}
\email{popov-isan@mail.ru}
\affiliation{Institute for Spectroscopy of Russian Academy of Sciences, Troitsk, Moscow 108840, Russia}

\begin{abstract}
The two-chain Frenkel-Kontorova model is applied for an analytical description of the energy and structure of the network of domain walls in bilayer graphene. Using this approach, the commensurate-incommensurate phase transition upon biaxial stretching of one of the graphene layers is considered. We demonstrate that formation of the equilateral triangular network of domain walls becomes energetically favourable above the critical relative biaxial elongation of the bottom layer of $3.0\cdot 10^{-3}$. It is shown that the optimal period of the triangular network of domain walls is inversely proportional to the difference between the biaxial elongation of the bottom layer and the critical elongation as long as it is much greater than the width of domain walls. Quantitative estimates of the contribution of a single dislocation node to the system energy and the period of the network of domain walls are obtained. Experimental measurements of the period could help to verify the energy of the fully incommensurate state (such as obtained by relative rotation of the layers) with respect to the commensurate one. 
\end{abstract}
\maketitle

\section{Introduction}
Stacking dislocations arising from variation in stacking of graphene layers were recently predicted for bilayer graphene \cite{Popov2011}. Since then, characteristic patterns consisting of commensurate domains separated by stacking dislocations, which play the role of incommensurate domain walls, have been observed in numerous experiments \cite{Alden2013,Butz2014,Lin2013,Yankowitz2014, Kisslinger2015,Jiang2016,Jiang2018}. It has been demonstrated that stacking dislocations affect electronic \cite{Hattendorf2013, San-Jose2014, Lalmi2014, Benameur2015, Koshino2013} and optical \cite{Gong2013} response of graphene and, similar to in-plane defects, they should be taken into account upon development of nanoelectronic and nanoelectromechanical devices.

In stacking dislocations, one of the layers in a bilayer is slightly stretched and the other one slightly compressed and/or there is a different shear strain in the layers so that the stacking changes from the one ground-state stacking to another one \cite{Popov2011, Lebedev2015, Lebedeva2016, Lebedeva2017, Lebedev2017}. The variation of the stacking is mostly localized in narrow strips with the width much smaller than the size of commensurate domains where the stacking is close to the ground-state one. These narrow strips are referred to as domain walls \cite{Alden2013,Butz2014,Lin2013,Yankowitz2014,Jiang2016,Jiang2018} or boundaries between commensurate domains \cite{Popov2011, Lebedev2015, Lebedeva2016, Lebedeva2017, Lebedev2017}. 

In previous theoretical works \cite{Popov2011, Lin2013, Butz2014, Lebedev2015, Lebedeva2016, Lebedev2017, Dai2016}, it was assumed that domain walls do not cross as long as the distance between them is large and can be treated as isolated. It was shown that formation of domain walls becomes energetically favourable above some critical uniaxial elongation of the bottom layer \cite{Popov2011, Lebedeva2016, Lebedev2017}. At small elongations, the layers are maintained commensurate by the interlayer interaction. However, at some critical elongation, the interlayer interaction energy can no longer compensate the large elastic energy. As a result, formation of the first domain wall takes place. The distance between adjacent domain walls decreases continuously as the elongation of the bottom layer is increased. Therefore, there is a commensurate-incommensurate phase transition of the second order, which can be characterized by the inverse distance between domain walls as the order parameter \cite{Popov2011,Pokrovsky1978,Chaikin1995}. So far such a phase transition has been studied for bilayer graphene  \cite{Popov2011, Lebedeva2016}, bilayer boron nitride \cite{Lebedev2015, Lebedeva2016} and graphene-boron nitride heterostructure \cite{Lebedev2017} only upon uniaxial stretching and crossing of domain walls was neglected.

\begin{figure}
	\centering
	\includegraphics[width=0.8\columnwidth]{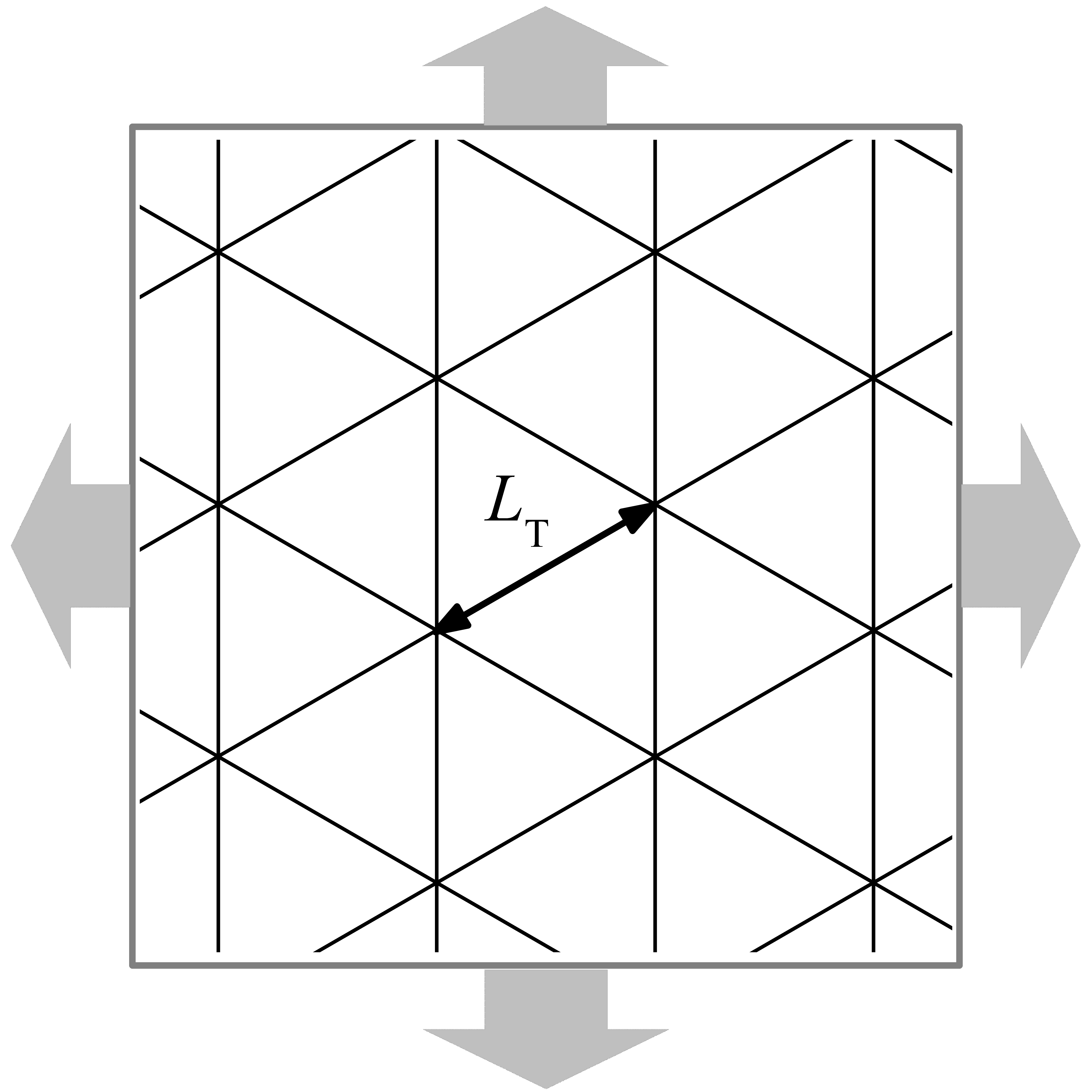}
	\caption{Schematic representation of the triangular network of domain walls (black lines) in bilayer graphene with a biaxially stretched bottom layer. The period $L_\mathrm{T}$ of the network is indicated. } 
	\label{fig:moire}
\end{figure}

Nevertheless, some experimental images of bilayer graphene obtained by transmission electron microscopy \cite{Alden2013, Kisslinger2015} and near-field infrared nanoscopy \cite{Jiang2016} revealed the presence of  triangular networks of intercrossed domain walls. Such networks are associated with global interlayer biaxial and rotation strains, i.e. different strains within the graphene layers and relative rotation of the layers \cite{Alden2013}. Therefore, deformation of one of the layers and/or its rotation with respect to the other should promote formation of networks of domain walls. In the present paper, we develop an analytical approach for description of networks of domain walls based on the two-chain Frenkel-Kontorova \cite{Popov2011,  Lebedev2015, Lebedeva2016,Bichoutskaia2006, Popov2009}. While diverse patterns of domain walls and loading conditions can be studied using this model, here we limit our consideration to the case when one graphene layer is biaxially stretched with respect to the other. The application of a tensile strain favors formation of tensile domain walls perpendicular to the Burgers vector \cite{Lebedeva2016, Lebedev2017}. Correspondingly, a regular network of tensile domain walls separating commensurate domains with the shape of equilateral triangles (Fig. \ref{fig:moire}) should arise upon application of a biaxial strain. We show that similar to uniaxial stretching, the commensurate-incommensurate phase transition takes place upon increasing the biaxial elongation of one of the graphene layers.  It should be noted that the commensurate-incommensurate phase transition considered here is very different from the crossover from the incommensurate state with commensurate domains separated by domain walls to the fully incommensurate state observed for graphene on boron nitride upon rotation of the graphene layer \cite{Woods2014}.

In the following, we review the theory of domain walls in bilayer graphene, develop the model for bilayer graphene with a biaxailly stretched bottom layer and apply it to estimate the critical biaxial elongation and the period of the network of domain walls. Finally we discuss the accuracy of our model and summarize our conclusions.

\section{Structure and energy of domain walls}
\begin{figure}
	\centering
	\includegraphics[width=\columnwidth]{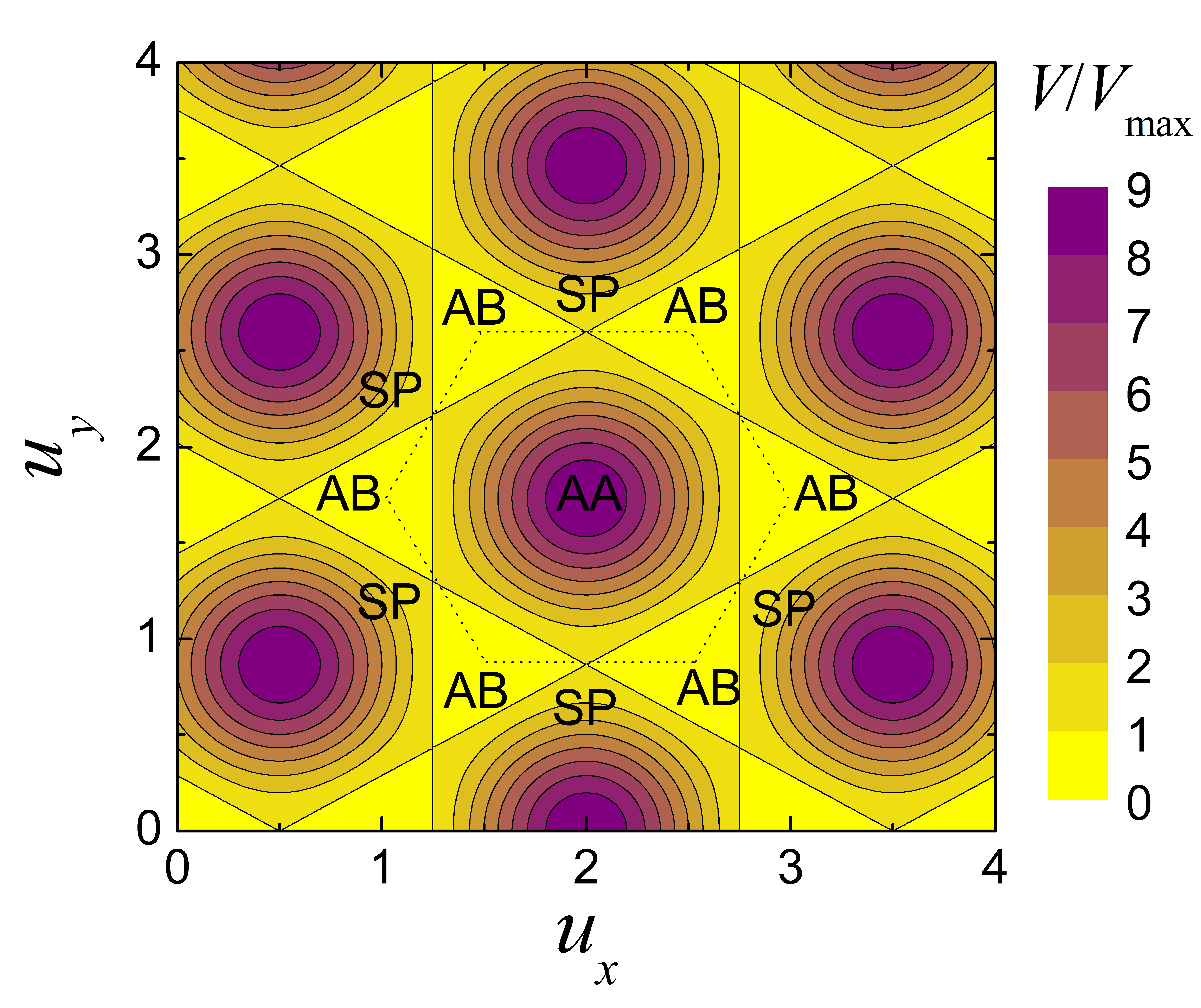}
	\caption{Interlayer interaction energy of bilayer graphene $V$ divided by the barrier $V_\mathrm{max}$ to relative sliding of the layers as a function of the relative displacements $u_x$ and $u_y$ of the graphene layers along the armchair and zigzag directions, respectively, in units of the bond length $l$. The interlayer interaction energy is computed according to  Eq. (\ref{eq_3}) and is given relative to the energy of the AB stacking. The AB, AA and SP stackings are indicated. The boundaries of a hexagon of the potential energy surface spanned by relative displacements of the layers within a single dislocation node are shown by dotted lines.} 
	\label{fig:pes}
\end{figure}

To understand the structure of domain walls in bilayer graphene and the reasons for formation of the network of domain walls, it is necessary to consider the potential surface of interlayer interaction energy of graphene layers, i.e. the dependence of the interlayer interaction energy on the relative in-plane displacement of the layers. 
This potential energy surface is determined by hexagonal symmetry of a single graphene layer  \cite{Popov2012, Kolmogorov2005, Reguzzoni2012, Aoki2007, Ershova2010, Lebedeva2011, Lebedeva2010, Lebedeva2011a} (Fig. \ref{fig:pes}). The minima of the potential energy surface are located in vertices of hexagons, the maxima in the centers of the hexagons and the barriers for transition between the minima in the middle of the sides of the hexagons. The maxima correspond to the AA stacking in which all atoms of one layer are on top of the atoms of the second one. The minima are observed for the AB stacking obtained from the AA stacking by the shift one of the layers by one bond length $l$ in any armchair direction. The barriers for transition between the minima correspond to the saddle-point (SP) stacking, which can be obtained from the AA stacking by the shift of one of the layers by $3l/2$ in any armchair direction. 

The structure and energy of domain walls depend only on changes of the interlayer interaction energy for different stackings and not on its absolute values. Therefore, in Fig. \ref{fig:pes} and below, we consider the interlayer interaction energy relative to the energy of the AB stacking, i.e. set it to zero at the minima of the potential energy surface.    

\begin{figure}
	\centering
	\includegraphics[width=0.8\columnwidth]{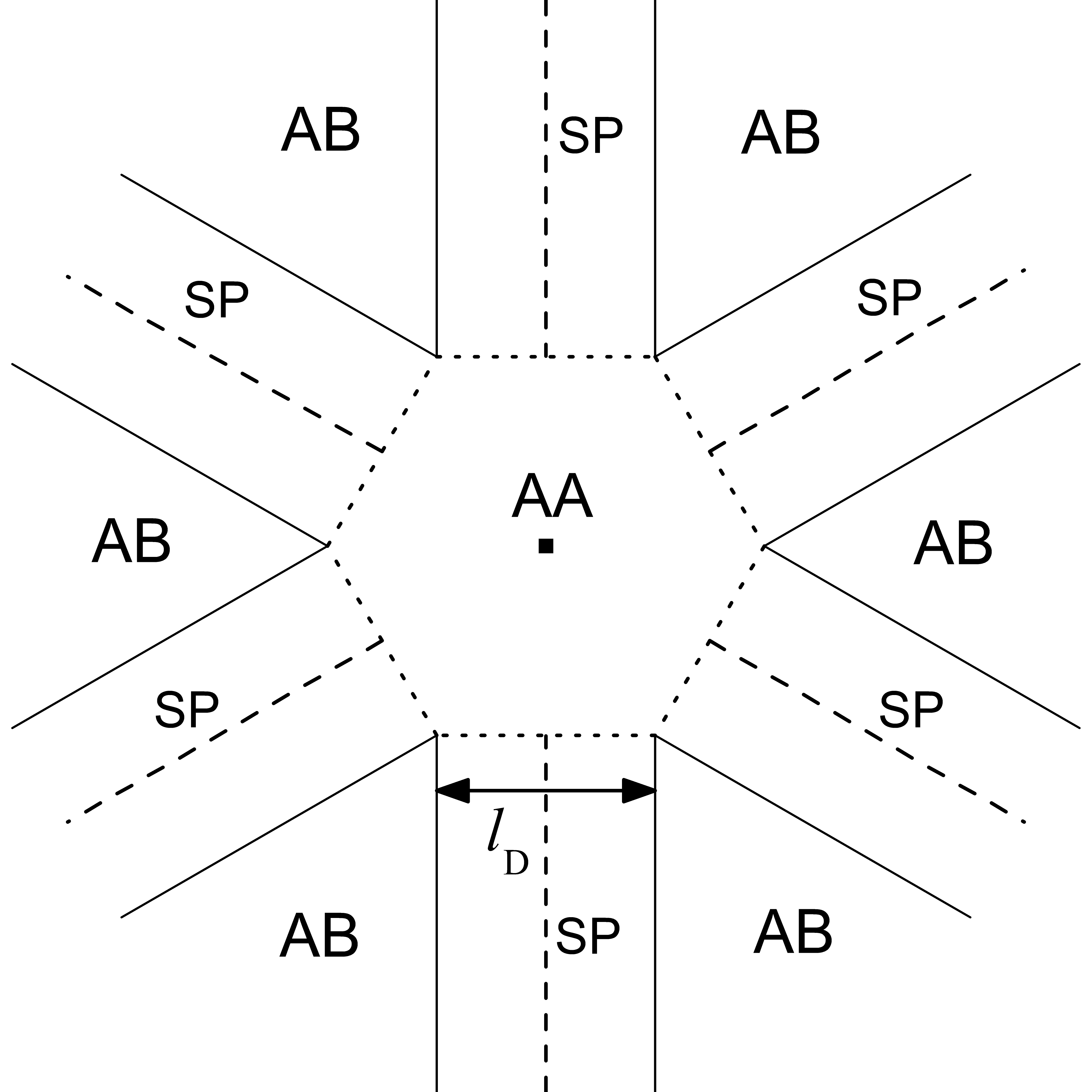}
	\caption{Schematic representation of a dislocation node in bilayer graphene with a biaxially stretched bottom layer. The boundaries between commensurate domains with the AB stacking and domain walls are shown using solid lines. The dislocation centers with the SP stacking are indicated using dashed lines. Dotted lines correspond to the boundaries of the dislocation node with the AA stacking in the center. The dislocation width $l_\mathrm{D}$ is indicated. } 
	\label{fig:node}
\end{figure}

Domain walls in bilayer graphene correspond to the relative displacement of the layers in the armchair direction through the SP stacking \cite{Popov2011, Lebedeva2016}, which is the minimum energy path between adjacent minima AB. The dislocations are partial since the Burgers vector $\vec{b}$ equal in magnitude to the bond length $l$ is smaller than the lattice constant $a_0 = l \sqrt{3}$. In the bilayer with a biaxially stretched bottom layer, a triangular network of domain walls should be formed with six domain walls corresponding to six armchair directions merging at each dislocation  node (Fig. \ref{fig:moire}). It is clear, therefore, that dislocation nodes correspond to the transition through the AA stacking (Fig. \ref{fig:node}).


\begin{figure}
	\centering
	\includegraphics[width=\columnwidth]{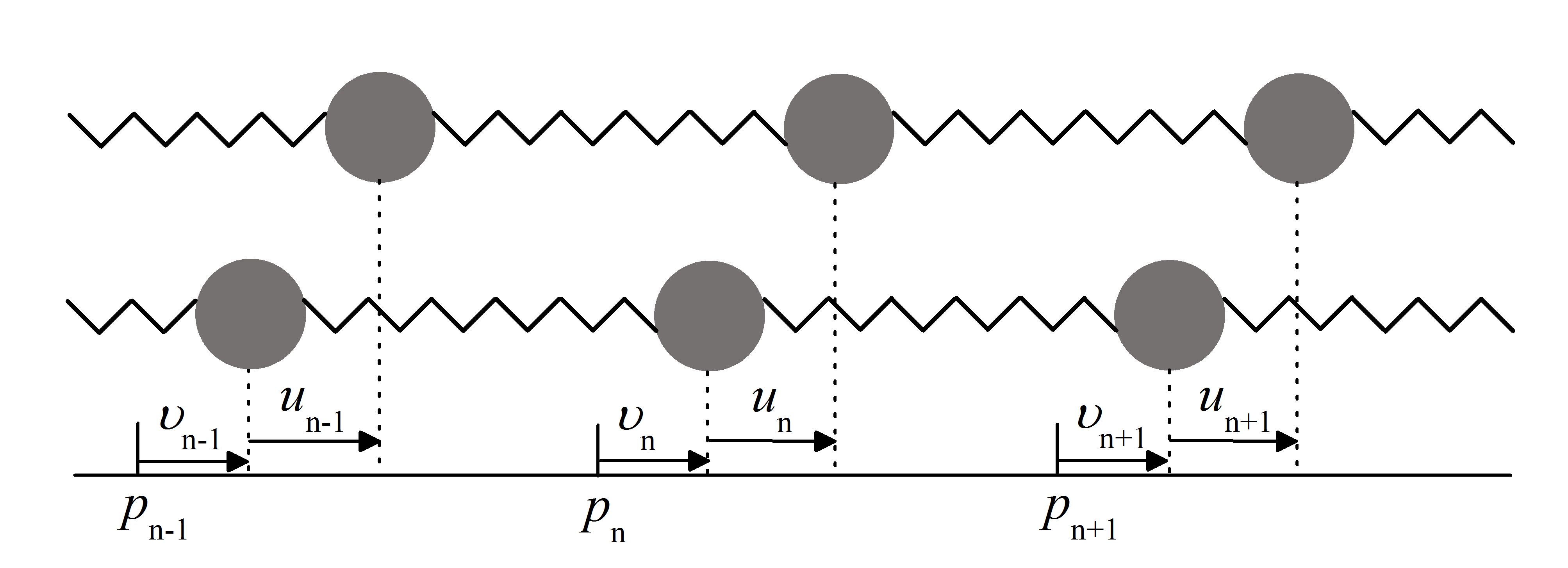}
	\caption{Scheme of the Frenkel-Kontorova model for two interacting chains of particles. } 
	\label{fig:FK}
\end{figure}

To study domain walls in bilayer graphene we use the two-chain Frenkel-Kontorova model  \cite{Popov2011,  Lebedev2015, Lebedeva2016,Bichoutskaia2006, Popov2009}, which takes into account structural relaxation of both of the layers and was applied previously to study domain walls in double-walled carbon nanotubes \cite{Bichoutskaia2006, Popov2009},  bilayer graphene  \cite{Popov2011,Lebedeva2016, Lebedeva2017},  bilayer boron nitride \cite{Lebedev2015} and graphene-boron nitride heterostructure \cite{Lebedev2017}. The out-of-plane buckling of graphene layers \cite{Butz2014,Lin2013} is neglected. This is justified when the bilayer is supported, for example, on misoriented graphene layers \cite{Alden2013}, boron nitride \cite{Yankowitz2014}, etc. In such cases, the two-chain Frenkel-Kontorova model parameterized on the basis of first-principles calculations \cite{Popov2011, Lebedeva2016} gives the width of domain walls in close agreement with the experimental data \cite{Alden2013, Lin2013, Yankowitz2014}. The account of out-of-plane buckling in suspended graphene can be performed using more advanced continuum models \cite{Dai2016} or atomistic approaches \cite{Butz2014,Lin2013}.

To describe a straight and isolated domain wall in free bilayer graphene within the two-chain Frenkel-Kontorova model, we consider two chains of particles connected by harmonic springs (Fig. \ref{fig:FK}). Each particle corresponds to a ribbon of graphene parallel to the domain wall, of unit length and of width equal to the bond length $l$. The number of particles in each chain $N\gg 1$. The elastic constant of the springs is $k_\mathrm{s}$. The equilibrium distance between particles in isolated chains is $l$. Then the elastic energy of the chains can be computed as 
\begin{equation} \label{eq_f1}
\begin{split}
W_\mathrm{el} = & \frac{k_\mathrm{s} l^2}{2} \sum\limits_{n=-N/2}^{N/2} \Big\{ \left(v_n-v_{n-1}\right)^2 \\& + \left(v_n+u_n-v_{n-1}-u_{n-1}\right)^2\Big\}, 
\end{split}
\end{equation}
where $v_n$ is the displacement of the $n$-th particle  in the chain corresponding to the bottom layer relative to the position $p_n=nl$ in the isolated chain and $u_n$ is the relative displacement of the $n$-th particles in the chains corresponding the upper and bottom layers. Both of these quantities are in the units of the bond length $l$.

The interaction energy of the chains is given by
\begin{equation} \label{eq_f2}
\begin{split}
W_\mathrm{int} &=  \sum\limits_{n=-N/2}^{N/2}V(u_n)l, 
\end{split}
\end{equation}
where $V(u)$ is the interlayer interaction energy of bilayer graphene per unit area relative to that in the commensurate state with the AB stacking. 

In the continuum approximation and in the limit $N\to \infty$, the total energy of the system, $W = W_\mathrm {el}+W_\mathrm{int}$, can be written as
\begin{equation} \label{eq_f3}
\begin{split}
W =  &\int\limits_{-\infty}^{+\infty} \Bigg\{ \frac{k_\mathrm{s} l^2}{2} \left(\frac{\mathrm{d}v}{\mathrm{d}n}\right)^2  + \frac{k_\mathrm{s} l^2}{2} \left(\frac{\mathrm{d}v}{\mathrm{d}n}+\frac{\mathrm{d}u}{\mathrm{d}n}\right)^2\\&+V(u)l\Bigg\}\mathrm{d}n.
\end{split}
\end{equation}

Introducing the variable $a = v + u/2$, the above equation can be presented in the form
\begin{equation} \label{eq_f4}
\begin{split}
W = & \int\limits_{-\infty}^{+\infty} \Bigg\{ k_\mathrm{s} l^2\left(\frac{\mathrm{d}a}{\mathrm{d}n}\right)^2 + \frac{k_\mathrm{s} l^2}{4} \left(\frac{\mathrm{d}u}{\mathrm{d}n}\right)^2\\&+V(u)l\Bigg\}\mathrm{d}n. 
\end{split}
\end{equation}

The energy described by Eq. (\ref{eq_f4}) is minimized when $\mathrm{d}a/\mathrm{d}n$ is set to zero and $0 \le u(n) \le 1$ corresponds to the relative displacement of atoms of the layers between two adjacent AB minima through the SP stacking (Figs. \ref{fig:pes} and \ref{fig:node}). In the present paper we limit our consideration to tensile domain walls. In this case, i.e. when the Burgers vector is perpendicular to the domain wall, $k_\mathrm{s}l=k/(1-\nu^2)$, where $k$ is the elastic constant of graphene under uniaxial stress (related to the Young's modulus $Y$ and the thickness of graphene layers $h$ as $k=Yh$) and $\nu$ is the Poisson's ratio. Then Eq. (\ref{eq_f4}) is reduced to
\begin{equation} \label{eq_1}
\begin{split}
\Delta W = & \int\limits_{-\infty}^{+\infty} \bigg\{\frac{k l^2}{4(1-\nu^2)} \left|\frac{\mathrm du}{\mathrm dx}\right|^2  + V(u)\bigg\}\mathrm{d}x,
\end{split}
\end{equation}
where $\Delta W=W$  is the energy associated with creation of a single domain wall per unit length and coordinate $x=nl$ corresponds to the direction perpendicular to the domain wall and parallel to the Burgers vector (i.e. along the armchair direction).

The relative displacement $u(x)$ that minimizes the formation energy of domain walls in Eq. (\ref{eq_1}) is determined by the Euler-Lagrange equation $\delta \Delta W/ \delta u=0$, which upon integration gives
\begin{equation} \label{eq_2}
\begin{split}
\frac{k l^2}{4(1-\nu^2)}\left|\frac{\mathrm du}{\mathrm dx}\right|^2 = V(u). 
\end{split}
\end{equation}

It was shown in the previous papers \cite{Popov2012,Reguzzoni2012,Lebedeva2011,Lebedeva2010,Lebedeva2011a} that the potential energy surface of bilayer graphene can be described with high accuracy by the expression containing only the first Fourier harmonics:
\begin{equation} \label{eq_3}
\begin{split}
V(u_x,u_y) = &V_0\Bigg(\frac{3}{2}+\cos\Big(2k_0u_x-\frac{2\pi}{3}\Big) \\
&-2\cos\Big(k_0u_x-\frac{\pi}{3}\Big)\cos\Big(k_0u_y\sqrt{3}\Big)\Bigg),
\end{split}
\end{equation} 
where $k_0 = 2\pi/3$ and $u_x$ and $u_y$ correspond to relative displacements of the layers in the armchair and zigzag directions, respectively, in units of the bond length $l$. Here, as mentioned above, the interlayer interaction energy is given relative to that in the AB stacking.  

For $u_y=0$ and $0 \le u_x=u(x) \le 1$, the dependence of the interlayer interaction energy along the straight line between two adjacent minima AB, which also corresponds to the minimum energy path, can be written as
\begin{equation} \label{eq_4}
\begin{split}
	V(u) = V_\mathrm{max}\left(2\cos{\left(k_0 u + \frac{2\pi}{3}\right)} +1\right)^2,
\end{split}
\end{equation}
where $V_\mathrm{max}=V_0/2$ is the barrier to relative sliding of graphene layers through the SP stacking. 

From Eqs. (\ref{eq_2}) and (\ref{eq_4}), it follows that the domain wall is described by a soliton:
\begin{equation} \label{eq_5}
\begin{split}
u(x) &= \frac{1}{2}+\\&\frac{3}{\pi}\arctan{\left[\frac{1}{\sqrt{3}}\tanh{\left(2\pi x \sqrt{\frac{V_\mathrm{max}(1-\nu^2)}{3kl^2}} \right)}\right]}.
\end{split}
\end{equation}

The dependence $u(x)$ has a nearly constant slope close to the center of the domain wall at $x=0$, which is roughly equal to $|\mathrm{d} u/\mathrm{d} x|_{x=0}$.
Therefore, the dislocation width corresponding to the characteristic width of domain walls can be defined as \cite{Popov2011, Lebedev2015, Lebedeva2016,Bichoutskaia2006, Popov2009}
\begin{equation} \label{eq_7}
	\begin{split}
		l_\mathrm{D} = \left|\frac{\mathrm{d} u}{\mathrm{d} x}\right|^{-1}_{x=0} =\frac{l}{2} \sqrt{\frac{k}{V_\mathrm{max}(1-\nu^2)}}.
	\end{split}
\end{equation}

The dislocation energy per unit length of domain walls based on Eqs. (\ref{eq_1}), (\ref{eq_2}) and (\ref{eq_4}) becomes
\begin{equation} \label{eq_6}
\begin{split}
\Delta W = \sqrt{\frac{k l^2}{(1-\nu^2)}} &\int_0^1 \sqrt{V(u)}\mathrm{d}u\\&=\sqrt{\frac{k l^2 V_\mathrm{max}}{(1-\nu^2)}}\left(\frac{3\sqrt{3}}{\pi}-1\right).
\end{split}
\end{equation}

The use of the continuum approximation in Eqs. (\ref{eq_f3}) -- (\ref{eq_1}) is justified if $l_\mathrm{D} \gg l$. In the present paper, we use the following parameters for the model obtained by density functional theory (DFT) calculations \cite{Lebedeva2016} using the the second version of the van der
Waals density functional (vdW-DF2) \cite{Lee2010}: $l = 1.430$~\AA, $k = 331 \pm 1$~J/m$^2$, $\nu = 0.174 \pm 0.002$ and $V_\mathrm{max} = 1.61$~meV/atom (in meV per atom of the upper/adsorbed layer). The latter value for the barrier is in agreement with the data deduced from shear mode frequencies of bilayer and few-layer graphene and graphite \cite{Popov2012, Lebedeva2016a} and within the range of other DFT results \cite{Kolmogorov2005,Reguzzoni2012,Aoki2007,Ershova2010,Lebedeva2011, Lebedeva2016a,Dion2004}. Using these parameters, we estimate that for tensile domain walls in bilayer graphene, $l_\mathrm{D} =13.4$ nm and this is indeed much greater than the bond length. This result is close to the corresponding experimental data of $10.1 \pm 1.4$ nm (Ref. \onlinecite{Alden2013}) and 11 nm (Ref. \onlinecite{Yankowitz2014}).

Let us now discuss the effect of uniaxial stretching of the bottom layer on the formation energy of domain walls. For that we consider relative elongations of the layers in the case when one domain wall is created in the system of finite length $L$ in the direction of the Burgers vector \cite{Popov2011}.  
According to the two-chain Frenkel-Kontorova model, atoms of both of the layers are displaced relative to their positions in the initial commensurate system upon creation of a domain wall \cite{Popov2011, Lebedev2015, Lebedeva2016, Bichoutskaia2006, Popov2009}. One domain wall corresponds to the relative displacement of the layers equal to the bond length $l$. Therefore, if the layers are free and composed of the same material, the resulting relative elongations of the bottom and upper layers with one domain wall are $\epsilon_0= l/(2L)$ and $-\epsilon_0$, respectively. This means that Eqs. (\ref{eq_1}) and (\ref{eq_6}) describe the energy of the finite bilayer with one domain wall and relative elongation of the bottom layer $\epsilon_0$ with respect to the commensurate system (the bilayer with no domain walls) with zero elongation. Analogously, these equations also describe the energy of the bilayer with one domain wall and relative elongation of the bottom layer $\epsilon$ with respect to the commensurate system with the relative elongation $\epsilon-\epsilon_0$. Thus, to calculate the energy of the bilayer with one domain wall  with respect to the commensurate system with the same relative elongation $\epsilon$ of the bottom layer, it is needed to substract from Eqs. (\ref{eq_1}) and (\ref{eq_6}) the elastic energy coming from the increase of the relative elongation of the commensurate bilayer from $\epsilon-\epsilon_0$ to $\epsilon$. While the formation energy of domain walls in the case when the relative elongation of the bottom layer is greater by the extra elongation $\epsilon_0$ compared to the commensurate bilayer (see Eqs. (\ref{eq_1}) and (\ref{eq_6})) is always positive, the formation energy of domain walls in the system with the fixed length of the bottom layer can be negative for sufficiently large $\epsilon$ (above the critical value) \cite{Popov2011, Lebedeva2016, Lebedev2017, Bichoutskaia2006, Popov2009}.

\section{Triangular network of domain walls}
For a biaxially stretched bottom layer, we consider the triangular network with domain walls in all zigzag directions (Fig. \ref{fig:moire}). We assume that commensurate domains are large compared to the width of domain walls so that the side of the triangles $L_\mathrm{T} \gg l_\mathrm{D}$. In this case, Eqs. (\ref{eq_1}) and (\ref{eq_6}) describe the energy of domain walls in the bilayer with the relative biaxial elongation of the bottom layer $\epsilon$ with respect to the commensurate system with the relative  biaxial elongation $\epsilon-\epsilon_0$, where $\epsilon_0=\sqrt{3}l/(2L_\mathrm{T})$. Correspondingly, the energy of the bilayer with the network of domain walls relative to the commensurate bilayer at the same relative elongation $\epsilon$ of the bottom layer can be written as
\begin{equation} \label{eq_8}
\begin{split}
\Delta W_\mathrm{T} = -\Delta W_\mathrm{el} + \Delta W_\mathrm{dw}+ \Delta W_\mathrm{dn},
\end{split}
\end{equation}
where $\Delta W_\mathrm{el}$ is the increase of the elastic energy of the commensurate bilayer due to the extra elongation $\epsilon_0$, $\Delta W_\mathrm{dw}$ and $\Delta W_\mathrm{dn}$ are the contributions of domain walls (edges of triangles) and dislocation nodes (vertices of triangles, see Fig. \ref{fig:node}) into the formation energy of the network of domain walls, respectively. Below we give these relative energies per unit area of the overlap between the layers of the bilayer. 

The increase of the elastic energy of the commensurate bilayer due to the extra elongation $\epsilon_0$ is simply given by
\begin{equation} \label{eq_9}
\begin{split}
 \Delta W_\mathrm{el} =&\frac{2k}{(1-\nu)}\left(\epsilon^2-\left(\epsilon-\epsilon_0\right)^2\right)\\&=\frac{2\sqrt{3}k\epsilon}{(1-\nu)}\frac{l}{L_\mathrm{T}}-\frac{3k}{2(1-\nu)}\left(\frac{l}{L_\mathrm{T}}\right)^2.
\end{split}
\end{equation}

The contribution of domain walls is $\Delta W_\mathrm{dw}=3L_\mathrm{T}\Delta W/(2S_\mathrm{T})$, where $\Delta W$ is the energy per unit length of domain walls given by Eq. (\ref{eq_6}) and $S_\mathrm{T}=\sqrt{3}L_\mathrm{T}^2/4$ is the area of one commensurate domain. Therefore,
\begin{equation} \label{eq_10}
\begin{split}
\Delta W_\mathrm{dw} = \frac{2l}{L_\mathrm{T}} \sqrt{\frac{3k V_\mathrm{max}}{(1-\nu^2)}}\left(\frac{3\sqrt{3}}{\pi}-1\right).
\end{split}
\end{equation}

The contribution of dislocation nodes should be positive and proportional to the density of nodes. It, therefore, should depend quadratically on the period of the network of domain walls $\Delta W_\mathrm{dn}\propto(l/L_\mathrm{T})^2$. 

From this assumption and Eqs. (\ref{eq_8})--(\ref{eq_10}), the expression for the energy of the bilayer with the  triangular network of domain walls relative to the commensurate system can be written as 
\begin{equation} \label{eq_13}
\begin{split}
\Delta W_\mathrm{T}= \frac{Al}{L_\mathrm{T}} +\frac{Bl^2}{L^2_\mathrm{T}},
\end{split}
\end{equation}
where 
\begin{equation} \label{eq_14}
\begin{split}
A=-2\sqrt{3} \left(\frac{k\epsilon}{(1-\nu)}-\sqrt{\frac{k V_\mathrm{max}}{(1-\nu^2)}}\left(\frac{3\sqrt{3}}{\pi}-1\right)\right)
\end{split}
\end{equation}
and $B$ is determined by the contribution $\Delta W_\mathrm{dn}$ of dislocation nodes  and the term of the elastic energy $\Delta W_\mathrm{el}$  quadratic in $l/L_\mathrm{T}$ (see Eq. (\ref{eq_9})). Both of these contributions to the total energy are positive and $B$ is positive.

It should be noted that limit $L_\mathrm{T}\to \infty$ corresponds to the commensurate system. Accordingly, the energy $\Delta W_\mathrm{T}$ of the bilayer with the network of domain walls with respect to the energy of the commensurate system given by Eq. (\ref{eq_13}) in this limit tends to zero.

The optimal period of the network of domain walls can be found from the conditions $\partial \Delta W_\mathrm{T}/\partial L_\mathrm{T}=0$ and $\partial^2 \Delta W_\mathrm{T}/\partial L_\mathrm{T}^2\ge0$. It follows from these conditions that when $A$ is positive, the optimal period of the network tends to infinity,  i.e. the ground state corresponds to the commensurate system. When $A$ is negative, the ground state corresponds to the bilayer with the network of domain walls characterized by the period
\begin{equation} \label{eq_16}
\begin{split}
L_0 = -\frac{2B}{A}l.
\end{split}
\end{equation}

From Eqs. (\ref{eq_13}) and (\ref{eq_16}), it can be checked that in the latter case, the relative energy of the bilayer with the optimal network of domain walls is given by 
\begin{equation} \label{eq_16_1}
\begin{split}
\Delta W_0 = -\frac{A^2}{4B}.
\end{split}
\end{equation}
Since $B$ is positive (see Eqs. (\ref{eq_15})), the structure with the triangular network of domain walls is indeed more energetically favourable than the commensurate system for negative $A$.

The critical relative biaxial elongation $\epsilon_\mathrm{c}$ above which the network of domain walls corresponds to the ground state of the system is determined by the condition $A=0$. Using Eq. (\ref{eq_14}), this gives
\begin{equation} \label{eq_17}
\begin{split}
 \epsilon_\mathrm{c}=\sqrt{\frac{V_\mathrm{max} (1-\nu)}{k(1+\nu)}}\left(\frac{3\sqrt{3}}{\pi}-1\right)=(1-\nu)\frac{\Delta W}{kl}.
\end{split}
\end{equation}
Note that this critical elongation $\epsilon_\mathrm{c}$ does not depend on the contribution  $\Delta W_\mathrm{dn}$ of dislocation nodes to the relative energy of the bilayer with the network of domain walls having a quadratic dependence on $l/L_\mathrm{T}$.

Using the parameters obtained by the DFT calculations \cite{Lebedeva2016} (see Sec. II), we estimate that the triangular network of domain walls corresponds to the ground state of bilayer graphene with a biaxially stretched bottom layer at relative elongations  $\epsilon > \epsilon_\mathrm{c}=3.0\cdot 10^{-3}$. For the bilayer with a uniaxially stretched layer, the critical elongation for becoming domain walls energetically favourable is given by $\epsilon_\mathrm{c0}=\Delta W/kl$ and is about $3.6\cdot 10^{-3}$ (Ref. \onlinecite{Lebedeva2016}). It is seen that the ratio of the critical biaxial and uniaxial elongations is determined by the Poisson's ratio: $\epsilon_\mathrm{c}/\epsilon_\mathrm{c0}=1-\nu$, i.e. the critical value for biaxial stretching is smaller than in the case of uniaxial stretching by only 20\%. This may seem surprising as the elastic energy is about twice greater upon introduction of the biaxial elongation compared to the same uniaxial elongation. The density of dislocations, however, is also different in two cases and the critical elongation turns out to be similar. 

From Eq. (\ref{eq_16}), it follows that above the critical elongation, the optimal period $L_0$ of the network of domain walls is proportional to $L_0\propto(\epsilon-\epsilon_\mathrm{c})^{-1}$.  The inverse quantity $L_0^{-1}$ can be considered as the order parameter of the phase transition. 
From Eq. (\ref{eq_16_1}), it can be concluded that  the relative energy of the bilayer with the optimal network of domain walls changes as $\Delta W_0 \propto -l^2/L_0^2 \propto-(\epsilon-\epsilon_\mathrm{c})^{2}$ above the critical elongation. This demonstrates that the commensurate-incommensurate phase transition in bilayer graphene with a biaxially stretched bottom layer is of the second order. 

The predicted dependence of the optimal period $L_0$ of the network of domain walls $L_0\propto(\epsilon-\epsilon_\mathrm{c})^{-1}$ is different from the logarithmic law  $L_0\propto\ln[(\epsilon-\epsilon_\mathrm{c})^{-1}]$ for the optimal separation between adjacent domain walls in the standard Frenkel-Kontorova model for a one-dimensional chain on a fixed substrate \cite{Chaikin1995}. This difference is related to the contributions to the energy of the system with domain walls that have a non-linear dependence on the inverse distance between the domain walls. In our case, such contributions come from the extra elongation of the bottom layer that is needed to fulfil the boundary conditions (see Eq. (\ref{eq_9})) and the presence of dislocation nodes. Both of these contributions are positive and depend quadratically the period of the network of domain walls. In the standard Frenkel-Kontorova model these contributions are absent. Instead the repulsive interaction between the domain walls that depends exponentially on their separation is taken into account \cite{Chaikin1995}. We assume that this exponential energy term can be neglected as compared to the quadratic terms for large periods of the network of domain walls, $L \gg l_\mathrm{D}$. 

Note that the quadratic contribution to the energy should be present in any two-layer system where there are boundary conditions for the substrate layer but it is not completely fixed (even in one-dimensional systems, bilayers with a uniaxially stretched bottom layer and no dislocation nodes \cite{Popov2011}, etc.). Therefore, in such systems, the separation between domain walls should also change inversely proportional to the difference between the elongation of the substrate layer and critical elongation upon the commensurate-incommensurate phase transition. 

To get quantitative estimates of the optimal period of the network of domain walls, we assume that the nodes have the shapes of hexagons with the side $l_\mathrm{D}$ corresponding to the width of domain walls (Fig. \ref{fig:node}). We also suppose that the layers within the nodes are uniformly stretched with the biaxial strain $\pm l/(2l_\mathrm{D})$. Accordingly, the relative displacement between the layers in these regions changes linearly and corresponds to the AB stacking at the vertices of the dislocation nodes and AA stacking at the centers (Fig. \ref{fig:node}). This means that the average interlayer interaction energy within the nodes is given by 
\begin{equation} \label{eq_11_0}
\begin{split}
V_\mathrm{in} = \frac{\int_\mathrm{hex} V(u_x,u_y)\mathrm{d}u_x\mathrm{d}u_y}{\int_\mathrm{hex} \mathrm{d}u_x\mathrm{d}u_y}=\frac{3}{2}V_0=3V_\mathrm{max},
\end{split}
\end{equation}
where the integration is performed over one hexagon on the potential energy surface with vertices at the AB stacking and center at the AA stacking  (Fig. \ref{fig:pes}) and we use Eq. (\ref{eq_3}) for the interlayer interaction energy $V(u_x,u_y)$. It is clear that $V_\mathrm{in}$ is equal simply to the average of the interlayer interaction energy over the whole potential energy surface. Therefore, in our model, the structure of the layers within the dislocation nodes can be referred to as fully incommensurate. A similar state is achieved when the layers are rotated with respect to each other by an arbitrary angle that does not correspond to any Moir\'e pattern \cite{Popov2012, Lebedeva2010, Lebedeva2011a} (and even in the structures corresponding to Moir\'e patterns, the average interlayer interaction energy is only slightly different from $V_\mathrm{in}$ as shown in Ref. \onlinecite{Yakobson}).

Considering that the layers are uniformly stretched within the nodes and the interlayer interaction energy is equal to its average over the potential energy surface, the contribution of dislocation nodes to the relative energy of the bilayer with domain walls can be written as 
\begin{equation} \label{eq_11}
\begin{split}
\Delta W_\mathrm{dn} = 3\left(\frac{l_\mathrm{D}}{L_\mathrm{T}}\right)^2\left(V_\mathrm{in} + \frac{k}{2(1-\nu)}\left(\frac{l}{l_\mathrm{D}}\right)^2\right).
\end{split}
\end{equation}

Taking into account Eqs. (\ref{eq_7}) and (\ref{eq_11_0}), this contribution can be presented as
\begin{equation} \label{eq_12}
\begin{split}
\Delta W_\mathrm{dn} = \left(\frac{l}{L_\mathrm{T}}\right)^2 \frac{3(5+2\nu)k}{4(1-\nu^2)}.
\end{split}
\end{equation}

Then the parameter $B$ in Eq. (\ref{eq_16}) takes the form
\begin{equation} \label{eq_15}
\begin{split}
B= \frac{3(7+4\nu)k}{4(1-\nu^2)}.
\end{split}
\end{equation}

\begin{figure}
	\centering
	\includegraphics[width=\columnwidth]{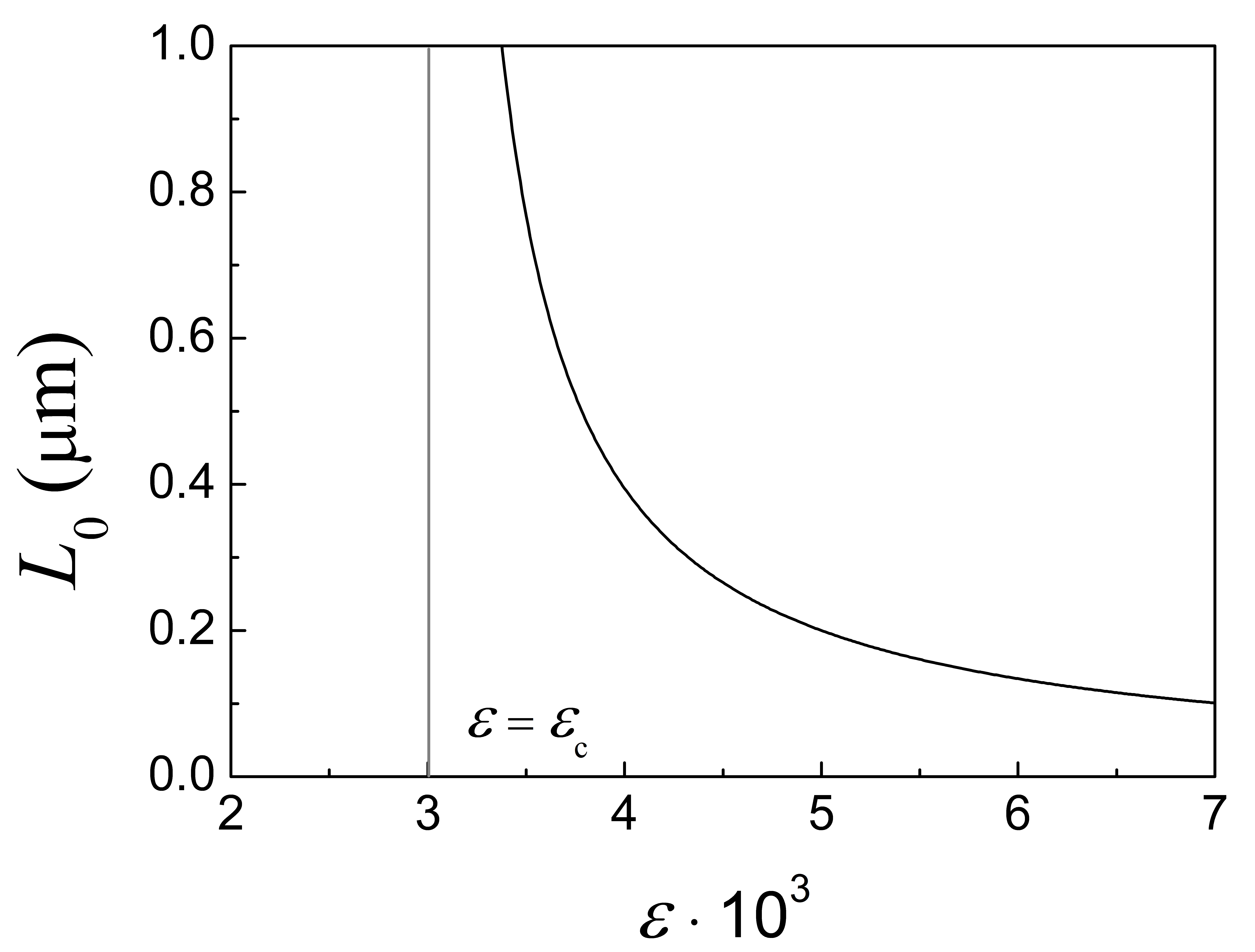}
	\caption{Estimated optimal period $L_0$ (in {$\mathrm\mu$}m) of the triangular network of domain walls in bilayer graphene as a function of the relative biaxial elongation $\epsilon$ of the bottom layer. The critical elongation $\epsilon_\mathrm{c}$ above which formation of such a network becomes energetically favourable is shown by the vertical  line.} 
	\label{fig:period}
\end{figure}

The dependence of the optimal period $L_0$ of the network of domain walls on the biaxial elongation estimated using Eq. (\ref{eq_16}) with the parameter $B$ from Eq. (\ref{eq_15})  is shown in Fig. \ref{fig:period}. The contribution of a single dislocation node to the relative energy of the bilayer with the triangular network of domain walls is $w_\mathrm{dn}=\Delta W_\mathrm{dn} \sqrt{3}L_\mathrm{T}^2/2=151$ eV.  This is the same as the contribution of a domain wall of the length of 0.14  {$\mathrm\mu$}m.

Our estimates are valid as long as $L_0 \gg l_\mathrm{D} =13.4$ nm. From  Eqs. (\ref{eq_7}), (\ref{eq_14}), (\ref{eq_16}) and (\ref{eq_15})  it follows that this condition is satisfied for 
\begin{equation} \label{eq_17}
\begin{split}
 \epsilon \ll \epsilon_\mathrm{c}+\frac{\sqrt{3}}{2}\left(7+4\nu\right)\sqrt{\frac{V_\mathrm{max} (1-\nu)}{k(1+\nu)}},
\end{split}
\end{equation}
which corresponds to $\epsilon \ll 3.3\cdot 10^{-2}$.

\section{Accuracy of the model}
Let us now discuss the accuracy of our model for dislocation nodes in which we assume that the layers are just uniformly stretched and fully incommensurate. A similar approximation can be also considered for domain walls, which formation energy is known from the exact analytical solution (see Eq. (\ref{eq_6})). For simple estimates, we can suppose that the layers are completely commensurate within the commensurate domains, while in a domain wall, the layers are uniformly stretched with the uniaxial tensile strain $\pm l/(2l_\mathrm{D})$ and the relative displacement between them changes linearly across the domain wall and lies on the straight line between two adjacent minima AB on the potential energy surface (Fig. \ref{fig:pes}). In this case, the formation energy of domain walls given by Eq. (\ref{eq_1}) can be written as
\begin{equation} \label{eq_18}
\begin{split}
\Delta W =  \frac{k l^2}{4(1-\nu^2)l_\mathrm{D}}  + V_\mathrm{av}l_\mathrm{D},
\end{split}
\end{equation}
where $V_\mathrm{av}$ is the average interlayer interaction energy along the line between two adjacent minima. 

From Eq. (\ref{eq_4}), it follows that
\begin{equation} \label{eq_19}
\begin{split}
V_\mathrm{av} =  \int_0^1 V(u)\mathrm{d}u= V_\mathrm{max}\left(3-\frac{9\sqrt{3}}{2\pi}\right).
\end{split}
\end{equation}

Optimization of the dislocation width $l_\mathrm{D}$ within this simple model using Eq. (\ref{eq_18}) and the condition $\partial \Delta W/\partial l_\mathrm{D}=0$, gives 
\begin{equation} \label{eq_21}
	\begin{split}
		l_\mathrm{D}=\frac{l}{2} \sqrt{\left(3-\frac{9\sqrt{3}}{2\pi}\right)^{-1}\frac{k}{V_\mathrm{max}(1-\nu^2)}}
	\end{split}
\end{equation}
and 
\begin{equation} \label{eq_22}
\begin{split}
\Delta W =\sqrt{\left(3-\frac{9\sqrt{3}}{2\pi}\right)\frac{k l^2 V_\mathrm{max}}{(1-\nu^2)}}.
\end{split}
\end{equation}
The latter value exceeds the result of the two-chain Frenkel-Kontorova model corresponding Eq.  (\ref{eq_6}) only by 10\%. 

The above value is the minimal error of the model in which the layers are uniformly stretched and incommensurate within domain walls. However, in the case of dislocation nodes formed by crossing domain walls, their size is determined by the dislocation width $l_\mathrm{D}$ from Eq.  (\ref{eq_7}). Using a similar assumption for domain walls, gives the formation energy 
\begin{equation} \label{eq_20}
\begin{split}
\Delta W =\frac{1}{2}\left(4-\frac{9\sqrt{3}}{2\pi}\right)\sqrt{\frac{k l^2 V_\mathrm{max}}{(1-\nu^2)}}.
\end{split}
\end{equation}
This is 16\% greater than the result given by Eq.  (\ref{eq_6}). Still this error is acceptable given that the barrier to relative sliding of the layers is not known with a high precision (the DFT values for the barrier lie in the range of 0.5 -- 2.1 meV/atom \cite{Kolmogorov2005,Reguzzoni2012,Aoki2007,Ershova2010,Lebedeva2011, Lebedeva2016a,Dion2004}, although the interval can be reduced if the experimental interlayer distance is used \cite{Lebedeva2016a}, and the estimates of the barrier from 
the experimental measurements of the shear mode frequencies \cite{Popov2012} and dislocation width \cite{Alden2013} correspond to 1.7 meV/atom and 2.4 meV/atom, respectively). It can be expected that our model used to analyze the triangular network of domain walls in bilayer graphene with a biaxially stretched bottom layer has a similar accuracy. To improve the accuracy of the model, a non-uniform strain distribution within dislocation nodes should be taken into account.

\section*{Conclusions and Discussion}
The two-chain Frenkel-Kontorova model has been applied to analyze the commensurate-incommensurate phase transition in bilayer graphene with a biaxially stretched bottom layer. It is found that the critical biaxial elongation above which the triangular network of domain walls corresponds to the ground state of the system does not depend on the structure and energetics of dislocation nodes. The critical biaxial elongation is estimated to be about $3.0\cdot 10^{-3}$ and its difference with the critical elongation in the case of the uniaxially stretched bottom layer is determined purely by the Poisson's ratio. It is also predicted that above the critical elongation, the optimal period of the network of domain walls changes inversely proportional to the difference between the biaxial elongation of the bottom layer and the critical elongation. Considering the inverse period of the network as the order parameter, it is proved that the commensurate-incommensurate phase transition is of the second order. 

To get quantitative estimates of the period of the triangular network of domain walls and energy of dislocation nodes, it is assumed that in the nodes, the layers are uniformly stretched and fully incommensurate. The estimated contribution of a single dislocation node to the total energy of the system is found to be about 150 eV. This quantity can be overestimated in our model of dislocation nodes by no more than 20\%, as demonstrated by the example of the formation energy of domain walls. 

In our estimates, we use the approximation of the potential energy surface of bilayer graphene by the first Fourier harmonics. This approximation has been justified by a number of DFT calculations \cite{Popov2012,Reguzzoni2012,Lebedeva2011,Lebedeva2010,Lebedeva2011a}. It was also applied to estimate the barrier $V_\mathrm{max}$ to relative sliding of graphene layers from measurements of the shear mode frequency of graphene layers in bilayer and few-layer graphene and graphite and the value of 1.7 meV/atom was deduced~\cite{Popov2012}, similar to the most reliable DFT results \cite{Lebedeva2016a}. The close value of 2.4 meV/atom was obtained from measurements of the dislocation width in few-layer graphene~\cite{Alden2013}. 
Experimental measurements of the period of the triangular network of domain walls, e.g., by  transmission electron microscopy \cite{Alden2013, Lin2013, Kisslinger2015}, scanning tunneling microscopy \cite{Yankowitz2014} or near-field infrared nanoscopy \cite{Jiang2016}, would allow to get an experimental estimate for the energy of transition of bilayer graphene from the commensurate state to the fully incommensurate one and to further check the accuracy of the approximation of the potential energy surface.

The approach for description of networks of domain walls in bilayer graphene developed in the present paper on the basis of the two-chain Frenkel-Kontorova model  has allowed us to get an analytical solution for the case of biaxial elongation of the bottom layer, i.e. when the commensurate domains have the shape of equilateral triangles and all the domain walls are tensile. 
In the experiments  \cite{Alden2013, Jiang2016, Kisslinger2015}, the triangular domains are not always equilateral and the domain walls are not equivalent. The corresponding  images can be obtained when strains applied to the bottom layer are not equal along different axes and/or there is a shear strain applied. Relative rotation of the layers and bending of the bilayer also affect the local structure of domain walls and the global structure of the network of domain walls. The current analytical approach based on the two-chain Frenkel-Kontorova model can be straightforwardly extended to describe formation of networks of domain walls for such types of the external load and they will be considered in subsequent papers.  

Other diverse patterns of domain walls have been observed experimentally including irregular triangular networks with curved domain walls  \cite{Alden2013, Kisslinger2015}, L-shape domain walls and closed-loop circles  \cite{Jiang2016}, etc. A non-uniform distribution of strains in the layers, presence of defects in graphene layers and substrate as well as existence of barriers to formation and transformation of domain walls \cite{Lebedeva2017} can be considered as possible reasons for formation of such patterns and require further investigation.

\section*{Acknowledgments}
AMP acknowledges the Russian Foundation for Basic Research (Grant 18-02-00985).

\bibliography{rsc}
\end{document}